\documentclass[11pt, conference]{IEEEtran}

\usepackage[T1]{fontenc} 
\usepackage[utf8]{inputenc} 
\usepackage{amsmath,cite,url}
\usepackage{graphicx}
\usepackage{color}
\usepackage{multirow}
\usepackage{subcaption}
\usepackage{enumitem}
\usepackage{dsfont}

\usepackage[bookmarks=false,hidelinks]{hyperref}

\begin{document}

\title{Cross-Modal Music-Video Recommendation: \\ A Study of Design Choices}



\author{\IEEEauthorblockN{}
\IEEEauthorblockA{}
\and
\IEEEauthorblockN{Laure Prétet}
\IEEEauthorblockA{LTCI, Télécom Paris \\ Bridge.audio, Paris, France}
\and
\IEEEauthorblockN{Gaël Richard}
\IEEEauthorblockA{LTCI, Télécom Paris \\ Institut Polytechnique de Paris, France}
\and
\IEEEauthorblockN{Geoffroy Peeters}
\IEEEauthorblockA{LTCI, Télécom Paris \\ Institut Polytechnique de Paris, France}}


\maketitle

\begin{abstract}

In this work, we study music/video cross-modal recommendation, i.e. recommending a music track for a video or vice versa.
We rely on a self-supervised learning paradigm to learn from a large amount of unlabelled data. 
We rely on a self-supervised learning paradigm to learn from a large amount of unlabelled data. 
More precisely, we jointly learn audio and video embeddings by using their co-occurrence in music-video clips. 
In this work, we build upon a recent video-music retrieval system (the VM-NET), which originally relies on an audio representation obtained by a set of statistics computed over handcrafted features.
We demonstrate here that using audio representation learning such as the audio embeddings provided by the pre-trained MuSimNet, OpenL3, MusicCNN or by AudioSet, largely improves recommendations. 
We also validate the use of the cross-modal triplet loss originally proposed in the VM-NET compared to the binary cross-entropy loss commonly used in self-supervised learning.
We perform all our experiments using the Music Video Dataset (MVD).

\end{abstract}


\section{Introduction}\label{sec:introduction}

Real-world data is inherently multimodal. 
When we observe our environment, we usually gather information using several senses at the same time.
Co-occurrence of events across our senses allows us to efficiently acquire a large amount of knowledge without the need of ground truth annotations.
Starting from this observation, an active research area has developed around multimodal learning paradigms \cite{Ngiam2011MultimodalLearning, Wang2016ARetrieval, Aytar2017SeeRepresentations, Muller2019Cross-ModalMethodologies}.
One of the most popular approaches 
associates audio data to corresponding video data to perform a variety of tasks \cite{Aytar2016SoundNet:Video, Owens2016AmbientLearning, Arandjelovic2017LookLearn, Hong2018CBVMR:Constraint, Schindler2019Multi-ModalAnalysis, Parekh2019WeaklyAnalysis}.
This association can be learned in a self-supervised way, thus requiring only a large dataset of unlabelled video clips, and no external annotation.
In this paper, we call "clip" a matching data pair composed of a silent video and an audio track, obtained from the same media and synchronized.
By learning to distinguish between matching and non-matching audio-video pairs, self-supervised systems can learn effective representations, or embeddings, of the audio and video.


Embeddings are non-linear projections of data in a high dimensional space tailored to solve a specific downstream task, such as classification.
In this study, we consider the intuitive downstream task of cross-modal recommendation.
Our problem is the following: given a query video clip, which music track from a database is the most suitable to serve as a soundtrack? 
And inversely, given a music track, which video from a database best illustrates its content?
Music recommendation for video has applications in automatic video editing \cite{Shah2014ADVISORRankings}, as well as professional music supervision \cite{Inskip2008MusicRetrieval}.
Video recommendation for music has applications in automatic MTV generation \cite{Liao2009MiningMTV} and music recommendation \cite{Zeng2018Audio-VisualCCA}.

By using the self-supervised paradigm, it is possible to solve the cross-modal recommendation task using only the content of the videos and music tracks.
Purely self-supervised systems do not exploit mood tags \cite{Sasaki2015AffectiveVideo,Li2019QueryRetrieval}, usage data or other prior information \cite{Shah2014ADVISORRankings}.
However, common audio-video self-supervised systems are computationally intensive to train (2 days on 16 GPUs in parallel for the L3-Net \cite{Arandjelovic2017LookLearn}).
One way to alleviate this issue is to leverage pretrained input features or to perform transfer learning.
In image and video processing, ImageNet features are commonly used as a starting point for other tasks \cite{Yue2015BeyondClassification, Balntas2018RelocNet:Nets, Wang2014LearningRanking}.
It has been shown that pretrained features associated to transfer learning or fine-tuning can match the performance of systems trained from scratch \cite{Oquab2014LearningNetworks}.
This allows to reduce the computational cost associated to the development of new systems, and to improve performances by leveraging a large diversity of datasets.
Reusing pretrained features can be critical in domains where access to training data is limited (e.g. due to copyright, or high cost of acquisition).
In music research, however, there is no widely accepted feature extractor that would be the equivalent of ImageNet features in image processing.
Several systems, both unimodal and multimodal, can play this role \cite{Pons2019Musicnn:Tagging, Cramer2019LOOKEMBEDDINGS, Gemmeke2017AudioEvents}.
We study here which of these audio embeddings is the most appropriate for the music-video recommendation task.


We build upon the VM-NET \cite{Hong2018CBVMR:Constraint}, a self-supervised network for music-video recommendation, and challenge some of its design choices.
We compare the performance of several open-source audio embeddings for this recommendation task, including MusiCNN, OpenL3 and AudioSet.
We show that audio features already pretrained on a cross-modal task \cite{Cramer2019LOOKEMBEDDINGS} perform best.
To our knowledge, this is the first time these embeddings are used in the context of music-video recommendation.

\textbf{Paper organization.}
In Section \ref{soa}, we review the literature about music-video embeddings.
In Section \ref{baseline}, we describe the reference system that we use as baseline \cite{Hong2018CBVMR:Constraint}.
In Section \ref{exps}, we challenge three design choices of our baseline system by changing the amount of training data, the training loss and the input audio features.
In Section \ref{eval}, the performances of these variants are evaluated on an independent test set.
Some conclusions are suggested in Section \ref{conclusion}.

\section{Related work}
\label{soa}

\subsection{Music-Video Embeddings}

Music-video embeddings refer to a joint representation between two different data modalities: music and video.
In general, music-video embeddings are computed via a pair of projection functions $(f_M, f_V)$.
$f_M$ takes as input a music audio excerpt $x_M$ and outputs an embedding vector $e_M=f_M(x_M)$. 
$f_V$ takes as input a video sample $x_V$ and outputs an embedding vector $e_V=f_V(x_V)$.
Both embedding vectors belong to the same high dimensional embedding space.
This allows to perform cross-modal retrieval, for example by computing Euclidean distances between two embedding vectors: $d_{a,b} = ||f_M(x_a) - f_V(x_b)||^2$.

$f_M$ and $f_V$ can be trained jointly to associate music and video samples according to some matching criteria.
In a broader sense this includes also systems designed for any type of sound and for images (single frame videos).
A popular design choice is to implement $f_M$ and $f_V$ as neural networks.
These can be trained with the help of another network that performs a fusion between both modalities, or using a specific loss that directly organizes the embedding vectors produced by $f_M$ and $f_V$.
The training of such embeddings can be done using two paradigms: supervised or self-supervised.

\vspace{0.2cm}
\textbf{Supervised approaches:}
In the case of supervised learning, the matching criterion that associates the audio and video modalities is deduced from additional sources of information. 
Typically, mood tags \cite{Shah2014ADVISORRankings}\cite{Zeng2018Audio-VisualCCA} or projections into the valence-arousal plane \cite{Sasaki2015AffectiveVideo} can be used to recommend musics and videos that have a similar emotional content.
The use of mood information accelerates the training, and allows the systems to reach promising retrieval performances. 
The matching criteria can be expressed as a score which is not necessarily binary, allowing nuances.
However, this restricts the systems to learn a certain type of audiovisual correspondence, and requires to collect additional information to train.


\vspace{0.2cm}
\textbf{Self-supervised approaches:}
In the case of self-supervised learning, a binary matching criterion is used. 
The learning objective is to classify pairs of audio and video samples between "matches" (i.e. both samples are extracted from the same clip and from the same time position) and "non-matches".

Early approaches have learned correlations between audio and video modalities \cite{Kuo2013BackgroundAnalysis} using multiple-type latent semantics analysis or dual-wing harmonium models \cite{Liao2009MiningMTV}.
Later, methods based on deep neural networks were proposed. 
With an appropriate loss function, neural networks can learn directly the correspondence between audio and video modalities. 
Aytar et. al. first experimented with a KL-divergence loss between the audio and video embedding vectors \cite{Aytar2016SoundNet:Video}.
Alternatively, the L3-Net proposed by Arandjelovic and Zisserman stacks a fully-connected fusion network after the two-branch embedding extraction network \cite{Arandjelovic2017LookLearn}. 
The whole system is trained with a binary cross-entropy loss, to predict whether both samples were extracted from the same video clip.
A variant was proposed by Owens and Efros, who use the cross-entropy loss to predict whether an audio and a video sample from the same clip were temporally shifted or not \cite{Owens2018Audio-VisualFeatures}.

These three last systems allow to leverage large non-annotated datasets of clips and the resulting embeddings were shown to perform well when reused for downstream tasks. 
However, their training is computationally expensive, and the embeddings are not specific to musical audio signals.

\subsection{Usages of Music-Video Embeddings}

\textbf{Use of audio embeddings as audio features for a downstream task:}
Once trained, music-video embeddings can be used as feature vectors for a variety of downstream tasks, including single-modal tasks.
For example, SoundNet~\cite{Aytar2016SoundNet:Video} is used for acoustic scene and object classification, while L3-Net~\cite{Arandjelovic2017LookLearn} and its open-source version OpenL3~\cite{Cramer2019LOOKEMBEDDINGS} are used for environmental sound classification.

Among the applications for cross-modal downstream tasks, works related to sounding object localization~\cite{Arandjelovic2018ObjectsSound} and audio-visual event localization~\cite{Tian2018Audio-visualVideos} are worth mentioning.
Owen and Efros also illustrate the performances of their system by audio-visual action recognition and on-screen (speaker visible on the screen) vs off-screen (not visible) audio source separation~\cite{Owens2018Audio-VisualFeatures}.

\vspace{0.2cm}
\textbf{Use of audio embeddings for cross-modal recommendation:}
Music-video embeddings can also be used for cross-modal recommendation or retrieval tasks. 
Examples of systems for music recommendation given video as input are \cite{Kuo2013BackgroundAnalysis,Shah2014ADVISORRankings,Sasaki2015AffectiveVideo,Shin2017MusicSimilarity,Li2019QueryRetrieval}.
In a symmetrical way, examples of systems for video recommendation given audio as input are \cite{Liao2009MiningMTV,Zeng2018Audio-VisualCCA}.

In this paper, we propose to evaluate our music-video embedding system on both music recommendation given video as input and video recommendation given music as input as Hong et. al.~\cite{Hong2018CBVMR:Constraint} did.

\section{Baseline system}
\label{baseline}

The baseline system that we consider in this paper is the "VM-NET" system proposed by Hong et. al.~\cite{Hong2018CBVMR:Constraint}\footnote{More precisely we re-implemented the VM-NET of \cite{Hong2018CBVMR:Constraint}.}. 
This system learns music-video embeddings in a self-supervised way (without the need for additional information) with the goal of performing cross-modal recommendation, hence music-to-video or video-to-music recommendation.
We briefly describe this system in the following (for more details see~\cite{Hong2018CBVMR:Constraint}).

\vspace{0.2cm}
\textbf{Input features:}
As opposed to other works, the input of the VM-NET are not the raw data (raw audio waveform or spectrogram, or images) but audio and image features extracted using a previous system.
Each video clip is represented by a pair of timeless vectors $(x_M, x_V)$ (one for each modality).

The audio input to the network, $x_M$, is a timeless feature vector of length 1,140 which represents the whole clip duration.
To construct this vector, handcrafted features (such as spectral centroid, MFCCs, and Chromas) are extracted at each time frame (1 frame$\simeq$21ms) and statistically aggregated (using mean, variance and max) over the whole clip duration (3.9 minutes on average).

The video input to the network, $x_V$, is a timeless feature vector of length 1,024 which also represents the whole clip duration.
To construct this vector, ImageNet features~\cite{Abu-El-Haija2016Youtube-8m:Benchmark} are computed for images of the videos sampled every second. The features are then statistically aggregated (using mean) over the whole clip duration.

\vspace{0.1cm}
\textbf{Network architecture:}
In the VM-NET, $x_M$ and $x_V$ feed two independent branches made of fully-connected layers with ReLu activations. 
The audio branch consists in 3 fully-connected layers of 2048, 1024, and 512 units respectively. 
The video branch consists in 2 fully-connected layers of 2048 and 512 units respectively.
Linear activations and batch normalization are applied at the end of each branch. Preliminary experiments validated the original architecture.

\vspace{0.1cm}
\textbf{Loss:}
The goal of the triplet loss \cite{Weinberger2009DistanceClassification} is to organize the projection $f$ of three samples in the embedding space such that the distance between an anchor $a$ and a positive sample $p$ is minimized, while the distance between the anchor $a$ and a negative sample $n$ is maximized.
The loss is defined as: $\mathcal{L}(a,p,n)= \max(||f(a)-f(p)||^2_2 - ||f(a)-f(n)||^2_2 + \alpha, 0)$, where $\alpha \in \mathds{R}_+^*$ is a margin parameter, and $f$ denotes the neural network embedding function.
In the VM-NET, the triplet loss is used not only to ensure that the music embeddings $e_M$ and video embeddings $e_V$ from the same clip are close but also that, for all clips, the distances between their music embeddings $e_M$ (resp. video embedding $e_V$) remains similar to the distance between their music features $x_M$ (resp. video features $x_V$).
This leads to the definition of an extended triplet loss $\mathcal{L}_{VMNET}$ with four constraints grouped into two types: inter-modal ranking constraints ($\mathcal{L}_{VM}, \mathcal{L}_{MV}$) and soft intra-modal structure constraints ($\mathcal{L}_{MM}$ and $\mathcal{L}_{VV}$).
These constraints are combined via a weighted sum: $\mathcal{L}_{VMNET}= \lambda_1 \mathcal{L}_{VM} + \lambda_2 \mathcal{L}_{MV} + \lambda_3 \mathcal{L}_{VV} + \lambda_4 \mathcal{L}_{MM}.$

\begin{figure}[ht]
\begin{subfigure}{\columnwidth}
  \centering
    \includegraphics[width=\columnwidth]{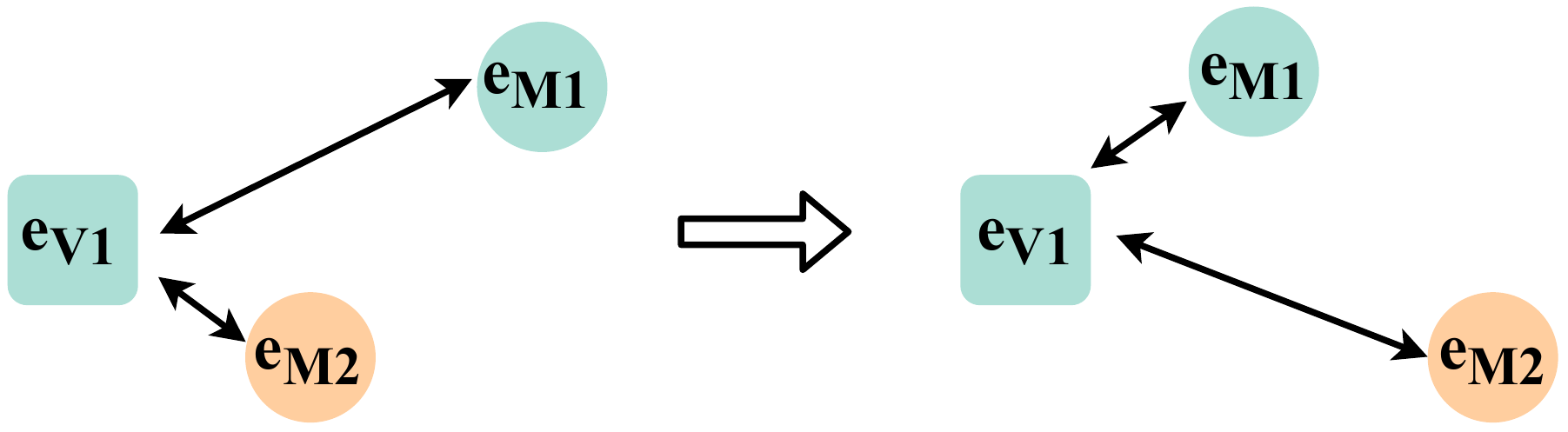}
    \caption{The $\mathcal{L}_{VM}$ inter-modal ranking constraint. Video $e_{V1}$ and music $e_{M1}$ taken from the same clip are brought together, while external music $e_{M2}$ is pulled away.}
    \label{fig:cross_modal_loss}
\end{subfigure}
\begin{subfigure}{\columnwidth}
  \centering
    \includegraphics[width=\columnwidth]{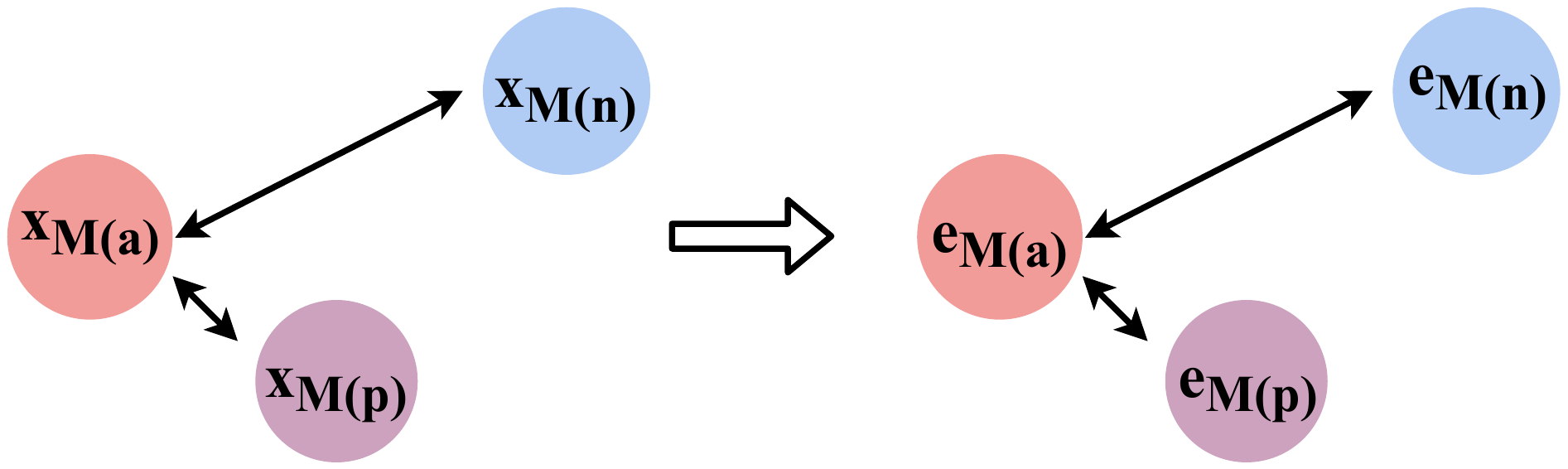}
    \caption{The $\mathcal{L}_{MM}$ music intra-modal structure constraint. Relative distances between music input features are preserved during training.}
    \label{fig:intra_modal_loss}
\end{subfigure}
\caption{The two types of constraints present in the $\mathcal{L}_{VMNET}$ loss. Squares represent the video modality, while circles represent the audio modality. Same colors represent the same clip.}
\label{fig:losses}
\end{figure}

For the \textit{inter-modal ranking constraints} $\mathcal{L}_{VM}$ and $\mathcal{L}_{MV}$, the triplets are defined as such: $a$ and $p$ are taken for the same clip, while $n$ is taken from a different clip (see Figure \ref{fig:cross_modal_loss}). 
The triplet loss is then computed in both directions: music to video $\mathcal{L}_{VM}$ with $(a,p,n) = (e_{V1}, e_{M1}, e_{M2})$ and video to music $\mathcal{L}_{MV}$ with $(a,p,n) = (e_{M1}, e_{V1}, e_{V2})$.
This ensures that each sample gets closer from its paired sample than from any other sample. 

The \textit{soft intra-modal structure constraints} are computed separately for the music ($\mathcal{L}_{MM}$) and for the video ($\mathcal{L}_{VV}$).
We only describe it for the music case (see Figure \ref{fig:intra_modal_loss}).
The process is the same for the video case.
The triplets are mined such that $a$, $p$ and $n$ are taken from three different clips but from the same modality (either music or video). 
For selecting $a$, $p$ and $n$ we use the values of the input features $x$ (not the values of the embeddings $e$). 
We select them such that the feature vectors $x_{M(a)}$ and $x_{M(p)}$ are closer (in terms of Euclidean distance) than $x_{M(a)}$ and $x_{M(n)}$.
The triplet loss is then applied to their corresponding embeddings: $(a,p,n) = (e_{M(a)}, e_{M(p)}, e_{M(n)})$ for $\mathcal{L}_{MM}$.
This ensures that the "modality-specific characteristics \textit{of the input features} are preserved even after the embedding process."
The process is the same for defining $\mathcal{L}_{VV}$ using $(a,p,n) = (e_{V(a)}, e_{V(p)}, e_{V(n)})$.

\vspace{0.2cm}
\textbf{Datasets:} We use two different datasets.

\textit{HIMV-50K dataset:}
The original VM-NET system is trained on a subset of the YouTube-8M dataset~\cite{Abu-El-Haija2016Youtube-8m:Benchmark}.
This subset corresponds to the clips annotated as "music video".
It is denoted as "HIMV-200K" in \cite{Hong2018CBVMR:Constraint} and consists of 205,000 video-music pairs.
It should be noted that while the audio of these clips always contains music, the video can be anything from professional promotional clips to amateur montages of still images.
From the list provided by the authors\footnote{\href{https://github.com/csehong/VM-NET/blob/master/data/Youtube_ID.txt}{https://github.com/csehong/VM-NET/blob/master/data/}}, we fetched all relevant YouTube IDs to re-create their dataset. 
However, due to country-specific restrictions, dead links and storage limitations, we only have access to 51,000 video-music pairs. 
We partition our dataset by leaving out 1,000 (randomly selected) pairs for validation, and we use the remaining 50,000 pairs for training.
We call it HIMV-50K in the following.

\textit{Music Video Dataset (MVD):}
While the previous dataset is large and diverse therefore suitable for training a neural network, the quality of the video part of the clips is very heterogeneous.
Therefore, to evaluate the performance of our system, we propose to use a cleaner dataset, the Music Video Dataset (MVD).
This is because evaluating the VM-NET on the noisy HIMV-50K dataset would make results difficult to interpret.
The MVD \cite{Schindler2017HarnessingRetrieval,Schindler2019Multi-ModalAnalysis} consists in 2,212 music video clips, manually curated. 
The music and video parts of the clips are of professional quality.
The average duration of each clip is 4 minutes.
We randomly selected $N=2,000$ of these clips to evaluate our systems.

\section{Proposed experiments}
\label{exps}

\subsection{Experiment 1: Size of the Training Data}

In the first experiment, we explore how critical the size of the training dataset is for music-video recommendation.
Intuitively, using pretrained visual features should alleviate the need for huge datasets.
To validate this, we train the VM-NET with three subsets of the HIMV-50K training dataset: 10k pairs, 30k pairs and 50k pairs. 
For the audio branch, we use the original handcrafted features.    

\subsection{Experiment 2: Triplet loss (TL) or Binary Cross-Entropy (BCE)}

As shown in the L3-Net paper\cite{Arandjelovic2017LookLearn}, it is possible to learn relevant audio-video embeddings with a binary cross-entropy loss.
In the second experiment, we then replace the $\mathcal{L}_{VMNET}$ triplet loss  by a binary cross-entropy loss.
To do so, we stack three fully-connected layers on top of the VM-NET with 1024, 128 and 1 unit respectively. 
The output activation is now a sigmoid.
The target of the network is set to 0 for non-matching audio-video pairs and 1 for matching pairs.
We therefore train the VM-NET for a binary classification task. 
For the audio branch, we still use the original handcrafted features.    

\subsection{Experiment 3: Input Audio Features}

\begin{figure}
    \centering
    \includegraphics[width=\columnwidth]{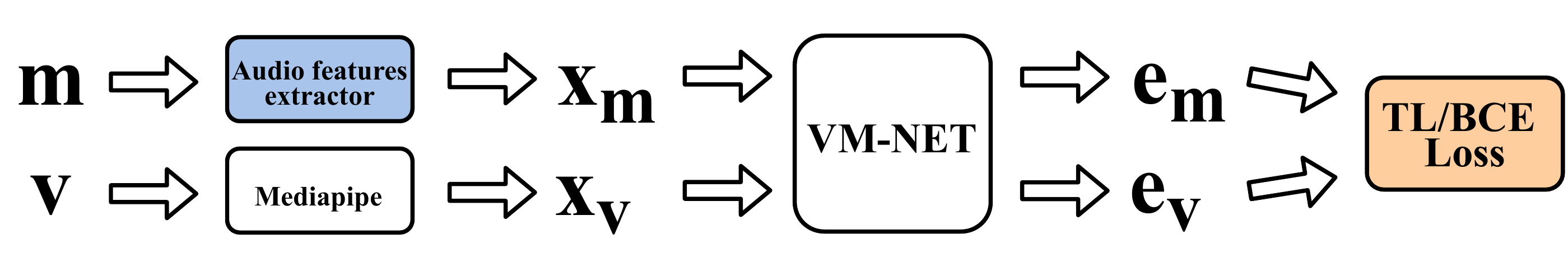}
    \caption{Experiment 2: Replacing the Triplet Loss by the BCE Loss (orange block). Experiment 3: Substituting handcrafted audio features for pretrained audio features (blue block).}
    \label{fig:audio_features}
\end{figure}

The VM-NET uses pretrained ImageNet features for its video branch and handcrafted audio features for its music branch.
This latter choice may be sub-optimal considering recent advances in feature learning using deep leaning.
We therefore gather four pretrained music embeddings and use them as audio feature extractors for our multimodal embedding training (See Figure \ref{fig:audio_features}).
We do not retrain them with the VM-NET, for a more efficient use of computation resources (we refer to the respective papers for more details on how each of them were trained).
We train each network with the $\mathcal{L}_{VMNET}$ triplet loss.

\textbf{MuSimNet:} We compute the embedding vectors using \texttt{MuSimNet}. This network has been proposed to estimate music similarity and trained using a triplet loss paradigm\cite{Pretet2020LearningLoss}.
For each track, we randomly select 24 segments of 12s duration and compute their corresponding embeddings. 
We then aggregate those over time using mean, variance and max, resulting in a single 384 dimensional feature vector for each music track.

\textbf{MusiCNN:} We compute the output of the penultimate layer of the music autotagging model MusiCNN \cite{Pons2019Musicnn:Tagging}. 
The open-source library \texttt{musicnn} was installed according to the instructions\footnote{\href{https://github.com/jordipons/musicnn}{https://github.com/jordipons/musicnn}}.
We compute the frame level features at a rate of 1Hz and we aggregate those over time using mean, variance and max, resulting in one single 600 dimensional feature vector for each music track.

\textbf{OpenL3:} We compute the music embeddings provided by the OpenL3 model.
OpenL3 has been trained to obtain embeddings usable for an audiovisual correspondence task.
This is a task similar to ours.
The open-source library \texttt{openl3} was installed according to the instructions\footnote{\href{https://github.com/marl/openl3}{https://github.com/marl/openl3}}.
We compute the frame level features at a rate of 1Hz and we aggregate those over time using mean, max and var, resulting in one single 18,432 dimensional feature vector for each music track. 
We use the system trained on the "music" dataset.

\textbf{AudioSet:} We downloaded the audio features provided by the YouTube-8M dataset\footnote{\href{https://research.google.com/youtube8m/download.html}{https://research.google.com/youtube8m/download.html}}.
These features are obtained from a model trained on an auto-tagging task on the YouTube-8M dataset.
That this is the same dataset as the one we use for training the VM-NET.
For the MVD, we compute these audio features with the open-source framework \texttt{mediapipe}.
We obtain one single 128 dimensional feature vector for each music track.


\section{Evaluation and results}
\label{eval}

\subsection{Training Details}
          
We reimplemented the VM-NET in Keras from the provided Tensorflow code\footnote{\href{https://github.com/csehong/VM-NET}{https://github.com/csehong/VM-NET}}.
After a preliminary validation experiment, we kept the original code and hyperparameters of the triplet loss $\mathcal{L}_{VMNET}$.
We train the VM-NET using only the HIMV-50K dataset.
We downloaded the YouTube-8M video features directly from the \href{https://research.google.com/youtube8m/download.html}{YouTube-8M webpage}. 
To compute the handcrafted audio features, we used the original VM-NET code, which makes use of the Librosa library \cite{McFee2015Librosa:Python}. 
In experiments 1 and 2, we use only these handcrafted audio features.
We use the ADAM optimizer with a learning rate of $10^{-6}$, a dropout scheme with probability 0.5 and a batch size of 1000.
For all TL systems, we use early stopping to prevent from overfitting.
For the BCE system, preliminary evaluations on the validation data showed that early stopping was detrimental for the performances. 
Instead, we train the VM-NET for 15,000 epochs.
All experiments are performed on a Nvidia GeForce GTX 1080 Ti GPU.
Running the evaluation script takes approximately 30 seconds for the TL systems, and 5 minutes for the BCE system.

\subsection{Evaluation Metrics}

Because of its higher quality, we only evaluate the VM-NET on the MVD.
Since the video features are not available for the MVD (they are only available for the HIMV-50K dataset), we compute those using \texttt{mediapipe}.
We evaluate our systems for two tasks:
\begin{itemize}[leftmargin=4mm, parsep=0cm, itemsep=0cm, topsep=0cm]
    \item \textit{Video to Music}: from a video query, retrieve the corresponding music.
    \item \textit{Music to video}: from a music query, retrieve the corresponding video.
\end{itemize}
In both cases, there is a single correct item to be retrieved.

To obtain the video recommendations given a music query, we first compute its embedding $e_M$ and compare it with all embeddings of the video from the test set $\{e_{V1}, ..., e_{VN}\}$.
When using the triplet loss, the comparison is done using the Euclidean distance (smaller distance means more similar).
When using the BCE loss, it is done using the sigmoid value (larger likelihood means more similar).
We then rank the video according to those.

For a given query $e_M$, if the corresponding video is in the top $k$ of the ranked list, we set the recall $R$ to 1, otherwise to 0.
For each query, $R$ is hence binary.
We repeat this operation using all music tracks of the test set as query.
To obtain the music recommendations given a video query, we swap the audio and video modalities.
The final metric is the average of $R@k$ over the 2,000 test clips, displayed as percentages. 
We use $k \in \{1,10,25\}$. The higher, the better.

\subsection{Results of the Experiment 1}




Table \ref{res1} shows the results of Experiment 1. 
The VM-NET trained on the Small dataset in 11 hours, on the Medium dataset in 14 hours and on the Large dataset in 18 hours.
Since the test set consists of 2,000 clips, we denote as "Chance" the recall expectancy of a random recommendation system.
We observe that the performances of the VM-NET improve when the size of the training dataset increases.
Although this is an expected result in most machine learning systems, it is not always obvious how critical the dataset size can be.
As performances do not seem to saturate in our experiments, we guess that the VM-NET could benefit from an even larger training set.

The difference between the results obtained for M$\rightarrow$V and V$\rightarrow$M may be explained by the distribution of the data in the embedding space. 
We computed the average dispersion of $e_v$ and $e_m$, and obtained 0.628 for music and 0.940 for video with the Large dataset.
Since the $e_v$ are more spread, it is easier to distinguish among them given a music query than it is to distinguish the $e_m$ given a video query.
The fact that the performance increases when increasing the training set size could however only be partly explained by this dispersion (which does not necessarily increase with the training set size).

\begin{table}
\caption{Results of Experiment 1: comparing different training-set sizes (test set: MVD). Conclusion: the more training data, the better the performances of the VM-NET.}
\label{res1}
\vspace{0.2cm}
\resizebox{\columnwidth}{!}{
\begin{tabular}{|c|c|c|c|c|c|c|}
    \hline
    \multirow{2}{*}{} & \multicolumn{3}{c|}{Music to Video} & \multicolumn{3}{c|}{Video to Music} \\ 
    \cline{2-7}
     & R@1 & R@10 & R@25 & R@1 & R@10 & R@25 \\ \hline
     Chance & 0.05 & 0.5 & 1.25 & 0.05 & 0.5 & 1.25 \\ \hline
     Small (10k) & 1.40 & 3.25 & 5.35 & 1.25 & 2.85 & 5.75 \\ 
     Medium (30k) & 1.60 & 5.30 & 8.90 & 1.25 & 4.85 & 9.35 \\ 
     Large (50k) & \textbf{2.05} & \textbf{7.85} & \textbf{13.30} & \textbf{1.85} & \textbf{7.00} & \textbf{12.90} \\ \hline
\end{tabular}
}
\end{table}

\begin{table}
\caption{Results of Experiment 2: comparing different training loss functions (test set: MVD, training set: Large). Conclusion: the Triplet Loss performs better than the BCE loss.}
\label{res2}
\vspace{0.2cm}
\resizebox{\columnwidth}{!}{
\begin{tabular}{|c|c|c|c|c|c|c|}
    \hline
    \multirow{2}{*}{} & \multicolumn{3}{c|}{Music to Video} & \multicolumn{3}{c|}{Video to Music} \\ 
    \cline{2-7}
    & R@1 & R@10 & R@25 & R@1 & R@10 & R@25 \\ \hline
     Chance & 0.05 & 0.5 & 1.25 & 0.05 & 0.5 & 1.25 \\ \hline
     TL & \textbf{2.05} & \textbf{7.85} & \textbf{13.30} & \textbf{1.85} & \textbf{7.00} & \textbf{12.90} \\
     BCE & 0.95 & 5.80 & 10.85 & 1.65 & 6.40 & 11.20 \\
     \hline
\end{tabular}
}
\end{table}

\begin{table} 
\caption{Results of Experiment 3: comparing different audio input features (test set: MVD, training set: Large, loss: TL). Conclusion: the handcrafted features are outperformed by pretrained embeddings such as AudioSet or OpenL3.}
\label{res3}
\vspace{0.2cm}
\resizebox{\columnwidth}{!}{
\begin{tabular}{|c|c|c|c|c|c|c|}
    \hline
    \multirow{2}{*}{} & \multicolumn{3}{c|}{Music to Video} & \multicolumn{3}{c|}{Video to Music} \\ 
    \cline{2-7}
    & R@1 & R@10 & R@25 & R@1 & R@10 & R@25 \\ \hline
     Chance & 0.05 & 0.5 & 1.25 & 0.05 & 0.5 & 1.25 \\ \hline
     Handcrafted & 2.05 & 7.85 & 13.30 & 1.85 & 7.00 & 12.90 \\ 
     MuSimNet & 1.30 & 7.05 & 14.75 & 0.80 & 7.20 & 13.20 \\ 
     MusiCNN & 1.60 & 9.30 & 18.55 & 1.65 & 8.45 & 16.90 \\ 
     AudioSet & 2.00 & 12.10 & 23.45 & 1.65 & \textbf{10.30} & \textbf{21.60} \\ 
     OpenL3 & \textbf{2.55} & \textbf{13.95} & \textbf{27.50} & \textbf{1.90} & 10.25 & 20.00 \\
     \hline
\end{tabular}
}
\end{table}


\begin{figure*}[h!]
\begin{subfigure}{\textwidth}
    \centering
    \includegraphics[width=\columnwidth]{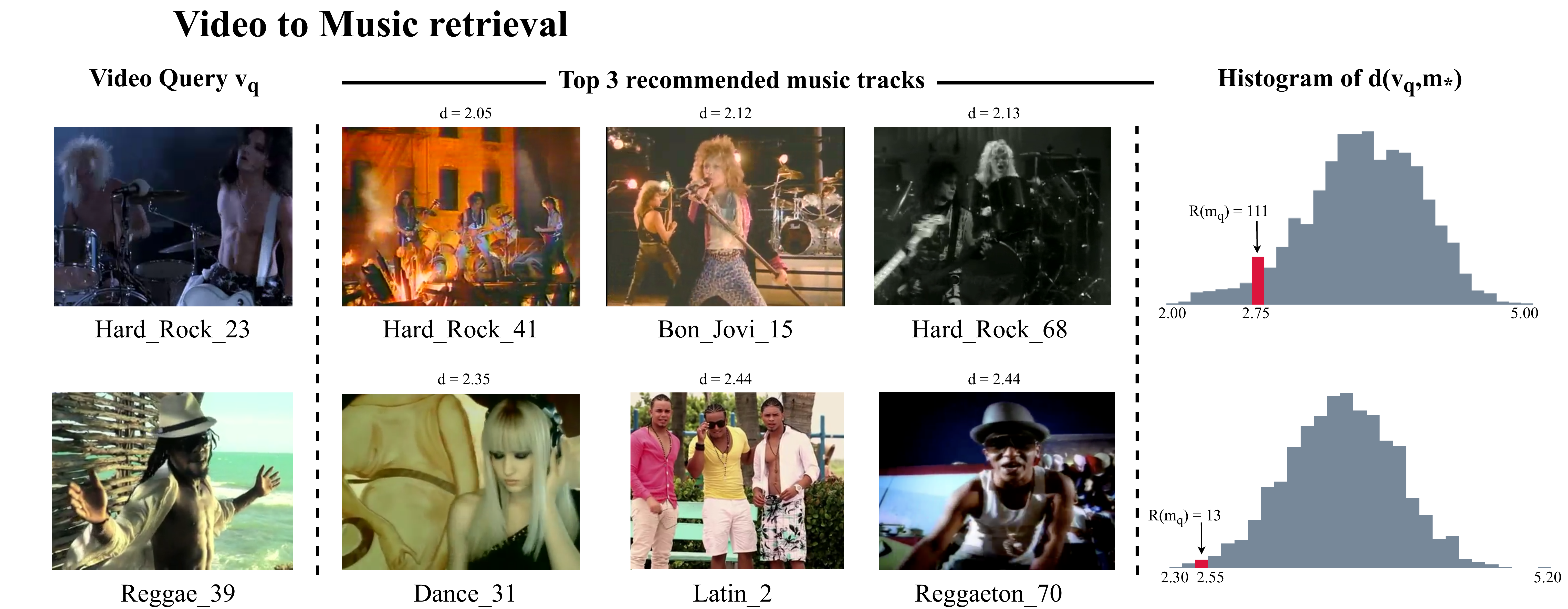}
    \label{fig:demo_vm}
\end{subfigure}
\begin{subfigure}{\textwidth}
    \centering
    \includegraphics[width=\columnwidth]{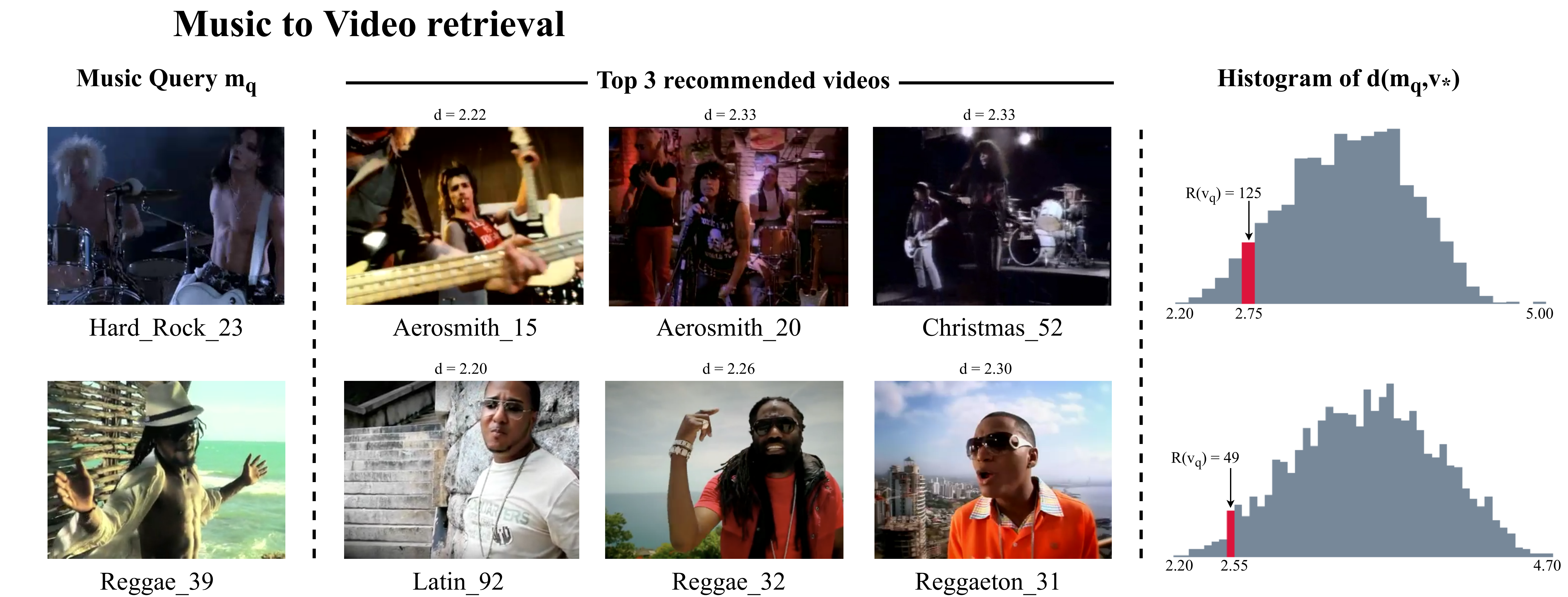}
    \label{fig:demo_mv}
\end{subfigure}
\caption{Qualitative results on the MVD for Video to Music and Music to Video recommendation tasks. All distances were multiplied by 1,000 for readability.}
\label{fig:demo}
\end{figure*}

\subsection{Results of the Experiment 2}


The VM-NET trained on with the triplet loss in 18 hours and with the BCE in 20 hours, so the training time of both systems is similar.
On Table \ref{res2}, we observe that replacing the $\mathcal{L}_{VMNET}$ loss by a the BCE seems to degrade the performances of the VM-NET.
We hypothesize that the triplet loss, that allows to perform metric learning, is more suited to our recommendation problem than the BCE, which trains a projection for a discriminative task.

\subsection{Results of the Experiment 3}





Table \ref{res3} shows the results of Experiment 3.
The two first rows show that the MuSimNet features perform comparably to the handcrafted features.
The fact that the MuSimNet features, which have been trained for a music similarity task, do not outperform the handcrafted ones can be explained by the fact that features that represent music similarity may not help representing audio to video similarity.
The MusiCNN features, which were trained on a music tagging task, perform a little better than the MuSimNet features.
Here, we assume that the VM-NET was able to take advantage of the music knowledge encapsulated in the MusiCNN features.

Finally, the two best performing embeddings are Audioset and OpenL3.
These two embeddings were trained on a much larger dataset than the MuSimNet and MusiCNN.
Note that the Audioset features were trained on a classification task on the YouTube-8M dataset, so the same content as for the video branch.
Note also that the OpenL3 features were trained on an audiovisual correspondence task, so a task similar to ours.
Additionally, the OpenL3 features were trained on music videos, which is closer to our application scenario.

The last row of Table \ref{res3} shows that when using OpenL3 features, the VM-NET achieves a R@25=27.5 for music-to-video.
This means that for more than 25\% of the music queries, the VM-NET was able to retrieve the exact corresponding video in its top 25 recommendations, out of 2,000 videos.
Remember that all test samples (queries and database) are taken from MVD, a different dataset than the one used for training.
Finally, we observe that the dimensionality $d$ of the audio feature vector does not seem to have any impact on the recommendation performance.
Indeed, the best performing features are the OpenL3 features ($d=18,432$) while the second best performing features are the Audioset features ($d=128$).
In comparison, the original handcrafted features have a dimension of 1,140.
However, $d$ seems to have an impact on the training time.
The VM-NET trained with the audio handcrafted features ($d=1,140$) in 18 hours, the MuSimNet features ($d=384$) in 18 hours as well, the MusiCNN features ($d=600$) in 19 hours, the AudioSet features ($d=128$) in 6 hours, and the OpenL3 features ($d=18,432$) in 42 hours.

\subsection{Qualitative Results}

To better understand the results, we provide in Figure~\ref{fig:demo} some examples of recommendations obtained by our best\footnote{We used the large training set, the triplet-loss $\mathcal{L}_{LVMNET}$ and the OpenL3 features.} system on the MVD.
We give 2 examples of Video query to Music [left part] and 2 examples of Music query to Video [right part].
We only display the first 3 recommendations provided by the system (from left to right).
As proposed by \cite{Hong2018CBVMR:Constraint}, we select one key frame (picture) of the clip to represent both the music track and the video.
On top of each picture, we indicate its Euclidean distance to the query (distances were multiplied by 1,000 for readability).
Below each picture, we provide the name of the clip as provided in the MVD.
To the right of the figure, we display the histogram of the distances between the query and all cross-modal samples.
We highlight in red the histogram bin corresponding to the ground truth sample (extracted from the same clip as the query), along with its rank in the list of recommendations.
Note that the exact matching sample of the query was retrieved in none of these examples, hence the recall at 3 for all these examples is zero.

As we see, while the R@3 is zero, the recommended musics (or videos) are from similar music genres than the video (or music) query (hard-rock for the first case, reggaeton/latin for the second) which still makes sense in terms of applications.
Such applications are recommendation systems for creative applications, e.g. music video editing or music supervision \cite{Inskip2008MusicRetrieval}.
In these cases, there is no ground truth music track to be retrieved.
This shows the limitations of the Recall metric, and a perceptual evaluation would provide complementary insights.

\section{Conclusion and perspectives}
\label{conclusion}

In this work, we studied music-video cross-modal recommendation.
We built upon a recent video-music retrieval system named VM-NET which originally relies on an audio representation obtained by a set of statistics computed over handcrafted features. 
We have demonstrated that using feature learning especially the audio embeddings provided by the pre-trained OpenL3 network allows to largely improve the obtained recommendations.
This may be explained by the fact that OpenL3 features were trained on a \textit{task similar to ours}, on a \textit{large dataset} of \textit{musical videos}.
Quoting \cite{Cramer2019LOOKEMBEDDINGS}, the "training data domain matches the downstream task".
The other audio features we considered (MuSimNet, MusiCNN and AudioSet) do not present these three characteristics simultaneously.

While using OpenL3 features improves the recommendations, there is still room for improvements such as first increasing the training set size.
However, it seems that for our specific task, downloading low-quality video  from YouTube will not help and that having access to professionally produced music-video clips is needed (as it is the case for the MVD dataset).
The fact that we did not fine-tune the audio feature extractors (MuSimNet, MusiCNN, AudioSet and OpenL3) along with the VM-NET also leaves room for potential improvement.
So far, the evaluation protocol requires that the system will recommend the exact same ground-truth video or audio from the pair, which is a severe metric.
A subjective human evaluation of the obtained recommendations would therefore help evaluating the quality of the recommendations.
Finally, the main restriction of the VM-NET is that it performs the recommendation at the full clip level, without considering any temporal variation within the clips.
Future studies will consider finer-grained temporal recommendations.


\bibliographystyle{splncs04}
\bibliography{references}

\begin{thebibliography}{10}
\providecommand{\url}[1]{\texttt{#1}}
\providecommand{\urlprefix}{URL }
\providecommand{\doi}[1]{https://doi.org/#1}

\bibitem{Abu-El-Haija2016Youtube-8m:Benchmark}
Abu-El-Haija, S., Kothari, N., Lee, J., Natsev, P., Toderici, G., Varadarajan,
  B., Vijayanarasimhan, S.: {Youtube-8m: A large-scale video classification
  benchmark}. arXiv preprint arXiv:1609.08675  (2016)

\bibitem{Arandjelovic2017LookLearn}
Arandjelovic, R., Zisserman, A.: {Look, Listen and Learn}. In: Proceedings of
  IEEE ICASSP (International Conference on Computer Vision). Venice, Italy
  (2017). \doi{10.1109/ICCV.2017.73}

\bibitem{Arandjelovic2018ObjectsSound}
Arandjelovi{\'{c}}, R., Zisserman, A.: {Objects that Sound}. In: Proceedings of
  ECCV (European Conference on Computer Vision). Munich, Germany (2018).
  \doi{10.1007/978-3-030-01246-5{\_}27}

\bibitem{Aytar2016SoundNet:Video}
Aytar, Y., Vondrick, C., Torralba, A.: {SoundNet: Learning Sound
  Representations from Unlabeled Video}. Advances in neural information
  processing systems pp. 892--900 (2016)

\bibitem{Aytar2017SeeRepresentations}
Aytar, Y., Vondrick, C., Torralba, A.: {See, Hear, and Read: Deep Aligned
  Representations}. arXiv preprint arXiv:1706.00932  (2017),
  \url{http://arxiv.org/abs/1706.00932}

\bibitem{Balntas2018RelocNet:Nets}
Balntas, V., Li, S., Prisacariu, V.: {RelocNet: Continuous Metric Learning
  Relocalisation using Neural Nets}. In: Proceedings of ECCV (European
  Conference on Computer Vision). Munich, Germany (2018)

\bibitem{Cramer2019LOOKEMBEDDINGS}
Cramer, J., Wu, H.H., Salamon, J., Bello, J.P.: {Look, Listen, and Learn More:
  Design Choices for Deep Audio Embeddings}. In: Proceedings of IEEE ICASSP
  (International Conference on Computer Vision). Brighton, UK (2019)

\bibitem{Gemmeke2017AudioEvents}
Gemmeke, J.F., Ellis, D.P.W., Freedman, D., Jansen, A., Lawrence, W.,
  Channing~Moore, R., Plakal, M., Ritter, M.: {Audio Set: An Ontology and
  Human-Labeled Dataset for Audio Events}. In: Proceedings of IEEE ICASSP
  (International Conference on Computer Vision). New Orleans, LA, USA (2017)

\bibitem{Hong2018CBVMR:Constraint}
Hong, S., Im, W., Yang, H.S.: {CBVMR: Content-based video-music retrieval using
  soft intra-modal structure constraint}. In: Proceedings of ACM ICMR
  (International Conference on Multimedia Retrieval). Yokohama, Japan (2018)

\bibitem{Inskip2008MusicRetrieval}
Inskip, C., Macfarlane, A., Rafferty, P.: {Music, Movies and Meaning:
  Communication in Film-makers’ Search for Pre-existing Music, and the
  Implications for Music Information Retrieval}. In: Proceedings of ISMIR
  (International Conference on Music Information Retrieval). Philadelphia, PA,
  USA (2008)

\bibitem{Kuo2013BackgroundAnalysis}
Kuo, F.F., Shan, M.K., Lee, S.Y.: {Background Music Recommendation for Video
  Based on Multimodal Latent Semantic Analysis}. In: Proceedings of ICME
  (International Conference on Multimedia and Expo). San Jose, CA, USA (2013)

\bibitem{Li2019QueryRetrieval}
Li, B., Kumar, A.: {Query by Video: Cross-Modal Music Retrieval}. In:
  Proceedings of ISMIR (International Conference on Music Information
  Retrieval). Delft, The Netherlands (2019)

\bibitem{Liao2009MiningMTV}
Liao, C., Wang, P.P., Zhang, Y.: {Mining Association Patterns between Music and
  Video Clips in Professional MTV}. In: Proceedings of MMM (International
  Conference on Multimedia Modeling). Sophia Antipolis, France (2009).
  \doi{10.1007/978-3-540-92892-8{\_}41}

\bibitem{McFee2015Librosa:Python}
McFee, B., Raffel, C., Liang, D., Ellis, D.P., McVicar, M., Battenberg, E.,
  Nieto, O.: {librosa: Audio and music signal analysis in python}. In:
  Proceedings of Python in Science. Austin, TX, USA (2015)

\bibitem{Muller2019Cross-ModalMethodologies}
M{\"{u}}ller, M., Arzt, A., Balke, S., Dorfer, M., Widmer, G.: {Cross-Modal
  Music Retrieval and Applications: An Overview of Key Methodologies}. In:
  Proceedings of IEEE ICASSP (International Conference on Computer Vision).
  Brighton, UK (2019). \doi{10.1109/MSP.2018.2868887}

\bibitem{Ngiam2011MultimodalLearning}
Ngiam, J., Khosla, A., Kim, M., Nam, J., Lee, H., Ng, A.Y.: {Multimodal Deep
  Learning}. In: Proceedings of ICML (International Conference on Machine
  Learning). Bellevue, WA, USA (2011)

\bibitem{Oquab2014LearningNetworks}
Oquab, M., Bottou, L., Laptev, I., Sivic, J.: {Learning and Transferring
  Mid-Level Image Representations using Convolutional Neural Networks}. In:
  Proceedings of IEEE CVPR (Conference on Computer Vision and Pattern
  Recognition). Columbus, OH, USA (2014)

\bibitem{Owens2018Audio-VisualFeatures}
Owens, A., Efros, A.A.: {Audio-Visual Scene Analysis with Self-Supervised
  Multisensory Features}. In: Proceedings of IEEE CVPR (Conference on Computer
  Vision and Pattern Recognition). Salt Lake City, UT, USA (2018)

\bibitem{Owens2016AmbientLearning}
Owens, A., Wu, J., McDermott, J.H., Freeman, W.T., Torralba, A.: {Ambient sound
  provides supervision for visual learning}. In: Proceedings of ECCV (European
  Conference on Computer Vision). Amsterdam, The Netherlands (2016).
  \doi{10.1007/978-3-319-46448-0{\_}48}

\bibitem{Parekh2019WeaklyAnalysis}
Parekh, S., Essid, S., Ozerov, A., Duong, N.Q.K., P{\'{e}}rez, P., Richard, G.:
  {Weakly Supervised Representation Learning for Audio-Visual Scene Analysis}.
  IEEE/ACM Transactions on Audio, Speech, and Language Processing  (2019)

\bibitem{Pons2019Musicnn:Tagging}
Pons, J., Serra, X.: {musicnn: Pre-trained convolutional neural networks for
  music audio tagging}. In: Late Breaking Demo, ISMIR (International Conference
  on Music Information Retrieval). Delft, The Netherlands (2019)

\bibitem{Pretet2020LearningLoss}
Pr{\'{e}}tet, L., Richard, G., Peeters, G.: {Learning to Rank Music Tracks
  Using Triplet Loss}. In: Proceedings of IEEE ICASSP (International Conference
  on Computer Vision). Barcelona, Spain (2020)

\bibitem{Sasaki2015AffectiveVideo}
Sasaki, S., Hirai, T., Ohya, H., Morishima, S.: {Affective Music Recommendation
  System Based on the Mood of Input Video}. In: Proceedings of MMM
  (International Conference on Multimedia Modeling). Sydney, Australia (2015)

\bibitem{Schindler2019Multi-ModalAnalysis}
Schindler, A.: {Multi-Modal Music Information Retrieval: Augmenting
  Audio-Analysis with Visual Computing for Improved Music Video Analysis}.
  Ph.D. thesis, Technische Universit{\"{a}}t Wien (2019)

\bibitem{Schindler2017HarnessingRetrieval}
Schindler, A., Rauber, A.: {Harnessing music-related visual stereotypes for
  music information retrieval}. ACM Transactions on Intelligent Systems and
  Technology  \textbf{8}(2), ~20 (2017). \doi{10.1145/2926719}

\bibitem{Shah2014ADVISORRankings}
Shah, R.R., Yu, Y., Zimmermann, R.: {ADVISOR - Personalized video soundtrack
  recommendation by late fusion with heuristic rankings}. In: Proceedings of
  ACM Multimedia. Orlando, FL, USA (2014). \doi{10.1145/2647868.2654919}

\bibitem{Shin2017MusicSimilarity}
Shin, K.H., Lee, I.K.: {Music synchronization with video using emotion
  similarity}. In: Proceedings of IEEE BigComp (International Conference on Big
  Data and Smart Computing). Jeju Island, South Korea (2017)

\bibitem{Tian2018Audio-visualVideos}
Tian, Y., Shi, J., Li, B., Duan, Z., Xu, C.: {Audio-visual event localization
  in unconstrained videos}. In: Proceedings of ECCV (European Conference on
  Computer Vision). Munich, Germany (2018)

\bibitem{Wang2014LearningRanking}
Wang, J., Song, Y., Leung, T., Rosenberg, C., Wang, J., Philbin, J., Chen, B.,
  Wu, Y.: {Learning Fine-grained Image Similarity with Deep Ranking}. In:
  Proceedings of IEEE CVPR (Conference on Computer Vision and Pattern
  Recognition). Columbus, OH, USA (2014)

\bibitem{Wang2016ARetrieval}
Wang, K., Yin, Q., Wang, W., Wu, S., Wang, L.: {A Comprehensive Survey on
  Cross-modal Retrieval}. arXiv preprint arXiv:1607.06215  (2016),
  \url{http://arxiv.org/abs/1607.06215}

\bibitem{Weinberger2009DistanceClassification}
Weinberger, K.Q., Saul, L.K.: {Distance metric learning for large margin
  nearest neighbor classification}. Journal of Machine Learning Research
  \textbf{10}(Feb),  207--244 (2009)

\bibitem{Yue2015BeyondClassification}
Yue, J., Ng, H., Hausknecht, M., Vijayanarasimhan, S., Vinyals, O., Monga, R.,
  Toderici, G.: {Beyond Short Snippets: Deep Networks for Video
  Classification}. In: Proceedings of IEEE CVPR (Conference on Computer Vision
  and Pattern Recognition). Boston, MA, USA (2015)

\bibitem{Zeng2018Audio-VisualCCA}
Zeng, D., Yu, Y., Oyama, K.: {Audio-Visual Embedding for Cross-Modal Music
  Video Retrieval through Supervised Deep CCA}. In: Proceedings of IEEE ISM
  (International Symposium on Multimedia). Taichung, Taiwan (2018).
  \doi{10.1109/ism.2018.00-21}

\end{thebibliography}

\end{document}